\documentclass[twocolumn,preprintnumbers,superscriptaddress,amsmath,amssymb]{revtex4-2}
\usepackage{float}
\usepackage{graphicx}% Include figure files
\usepackage{dcolumn}% Align table columns on decimal point
\usepackage{bm}% bold math
\usepackage{epstopdf}
\usepackage{xcolor}
\usepackage{textcmds}
\usepackage{textcomp}
\usepackage{soul}% \st
\usepackage[utf8]{inputenc}
\usepackage{csquotes}
\usepackage{amsbsy}
\usepackage{mathrsfs} % script-like, curvy letters.
\usepackage{amsmath}
\usepackage{bigints} % Big Integration
\usepackage{amsthm,amssymb} % Theorem Proof
\usepackage[english]{babel}
\usepackage[shortlabels]{enumitem}
\usepackage{float}
\usepackage{mathrsfs,bigints,mathtools,dsfont}
\usepackage[colorlinks=true,linkcolor=blue,citecolor=blue]{hyperref}%
\usepackage[toc]{appendix}

\begin{document}

\title{A strategy to control synchronized dynamics in swarmalator systems}

\author{Gourab Kumar Sar} \thanks{These authors contributed equally to this work} \affiliation{Physics and Applied Mathematics Unit, Indian Statistical Institute, 203 B. T. Road, Kolkata 700108, India}
 
\author{Md Sayeed Anwar} \thanks{These authors contributed equally to this work} \affiliation{Physics and Applied Mathematics Unit, Indian Statistical Institute, 203 B. T. Road, Kolkata 700108, India} 

\author{Martin Moriam\'e}\affiliation{Department of Mathematics and Namur Institute for Complex Systems, naXys, University of Namur, 2 rue Grafé, Namur B5000, Belgium} 

\author{Dibakar Ghosh}\affiliation{Physics and Applied Mathematics Unit, Indian Statistical Institute, 203 B. T. Road, Kolkata 700108, India}

\author{Timoteo Carletti}\email{timoteo.carletti@unamur.be}\affiliation{Department of Mathematics and Namur Institute for Complex Systems, naXys, University of Namur, 2 rue Grafé, Namur B5000, Belgium}

\begin{abstract}
Synchronization forms the basis of many coordination phenomena in natural systems, enabling them to function cohesively and support their fundamental operations. However, there are scenarios where synchronization disrupts a system’s proper functioning, necessitating mechanisms to control or suppress it. While several methods exist for controlling synchronization in non-spatially embedded oscillators, to the best of our knowledge no such strategies have been developed for swarmalators (oscillators that simultaneously move in space and synchronize in time). In this work, we address this gap by introducing a novel control strategy based on Hamiltonian control theory to suppress synchronization in a system of swarmalators confined to a one-dimensional space. The numerical investigations we performed, demonstrate that the proposed control strategy effectively suppresses synchronized dynamics within the swarmalator population. We studied the impact of the number of controlled swarmalators as well as the strength of the control term, in its original form and in a simplified one.
\end{abstract}

\maketitle

\section{Introduction}
Swarmalators, a portmanteau of ``swarm" and ``oscillators", represent a fascinating class of dynamical systems where agents exhibit both swarming behaviors and oscillatory synchronization~\cite{o2017oscillators}. These systems extend the conventional framework of swarming by incorporating (internal) phase dynamics, resulting in complex behaviors where spatial interactions are coupled with the synchronization of internal states~\cite{sar2022dynamics,o2019review}. This hybrid nature makes swarmalators a rich subject of study, as they offer a bridge between traditional swarm systems, such as those found in biological or robotic swarms, and synchronization phenomena seen in networks of oscillators.

In recent years, swarmalators have garnered significant attention for their ability to exhibit collective synchronized phenomena \cite{sar2022swarmalators,ghosh2024amplitude,anwar2024forced,o2024solvable,sar2023pinning,lizarraga2023synchronization,lizarraga2020synchronization} in systems where swarming and synchronization co-occur, with applications spanning biological microswimmers \cite{tamm1975role,belovs2017synchronized,yang2008cooperation,riedel2005self}, robotic swarms \cite{barcis2020sandsbots,monaco2020cognitive,rinner2021multidrone}, magnetic domain walls \cite{hrabec2018velocity,haltz2021domain}, and beyond \cite{yan2012linking,manna2021chemical,bricard2015emergent,zhang2020reconfigurable}. In swarmalator systems, while synchronized states — where, i.e., all units try to move and act in unison — are often desirable as phase and spatial interaction strengths increase, asynchronous behaviors, where units act differently simultaneously, can also play an equally important role in certain contexts. For example, in nature, flocks of birds or schools of fish often need flexibility to respond to predators or environmental changes. If they were too synchronized, their collective behavior might become rigid and less adaptable. Similarly, in technology, groups of robots might need to perform different tasks simultaneously instead of all doing the same thing. Asynchronous behavior allows these systems to stay flexible, adapt to changing conditions, explore the environment and solve complex problems more effectively.

In this paper, we thus focus on controlling the synchronization dynamics of swarmalators. By introducing external inputs or modifying interaction rules, we aim to steer the collective behavior towards desired (de)synchronied states, allowing for enhanced control over both the spatial arrangement and the phase coupling. Let us emphasize that, because of the very nature of swarmalators where spatial and internal dynamics are coupled, the proposed control strategy can achieve its goal by only acting on the internal phases. The ability to guide synchronization patterns has profound implications for real-world applications, such as sensor networks, coordinated robotics, and autonomous vehicles. For instance, controlling synchronization can optimize energy consumption, improve communication efficiency, and enhance task performance in multi-agent systems, e.g., exploration.

The control of collective synchronization to mitigate undesired effects has been extensively studied for non-spatial oscillator systems \cite{gjata2017using,asllani2018minimally,moriamelucascarletti2024}. For instance, researchers have proposed desynchronization strategies for Kuramoto oscillators (a prototype oscillator model for phase synchronization) by using Hamiltonian control theory \cite{vittot2004perturbation,chandre2005channeling}. Despite the dissipative nature of the Kuramoto model, Witthaut and Timme \cite{witthaut2014kuramoto} demonstrated that it can be embedded within a Hamiltonian system possessing an invariant torus, upon which the dynamics  mirror those of the Kuramoto model. Phase synchronization occurs only when this torus becomes unstable, making the control of its stability the core strategy for regulating synchronized dynamics \cite{witthaut2014kuramoto,gjata2017using}.

Here, we aim to extend the Hamiltonian control strategy to spatially embedded oscillator systems (swarmalators) for regulating their synchronized behavior. In this effort, we focus on a population of swarmalators whose movements are confined to a one-dimensional (1D) periodic domain. This simplified 1D swarmalator model is considered because it shares structural similarities with the coupled Kuramoto model and can also replicate the dynamics of more complex swarmalator systems \cite{o2022collective,anwar2024collective}. Building on the approach proposed in Ref.\cite{witthaut2014kuramoto}, we therefore introduce a Hamiltonian system that embeds the 1D swarmalator model within an invariant torus. Using this Hamiltonian framework, we propose a theoretical control term designed to suppress synchronized dynamics in the swarmalator population, by acting on the phases only. This approach generalizes the method presented in Ref.\cite{gjata2017using,asllani2018minimally} to spatially embedded systems.
We demonstrate that the proposed control term is highly effective in preventing the formation of synchronized collective states among the swarmalators. Notably, this control mechanism operates only when necessary: even if always present, it remains ineffective when the swarmalators are in a desynchronized state, allowing the controlled and uncontrolled systems to exhibit similar dynamics. However, the control term becomes effective precisely when the swarmalators attempt to synchronize their motion. Furthermore, we show that the synchronized dynamics can be effectively suppressed even when the control term is applied to only a fraction of the swarmalator population.

The rest of the paper is organized as follows. Section \ref{sec2} provides a brief review of the dynamics of the 1D swarmalator model. In Sec. \ref{sec3}, we introduce the Hamiltonian system for the swarmalator model. The derivation of the necessary control term is presented in Sec. \ref{sec:hamctrl}. Finally, in Sec. \ref{sec:numres}, we present numerical investigations demonstrating the effectiveness of the proposed control strategy.

%Additionally, this control extends beyond technical applications, finding relevance in biological systems such as collective motion in animal groups, synchronization of neural circuits, and even the formation of patterns in bacterial colonies. By exploring the various methods of synchronization control, this work contributes to a growing body of research that seeks to harness the potential of swarmalators in both artificial and natural systems.
%
%The study of swarmalator synchronization offers valuable insights into several critical applications in swarm control. For example:
%
%Robotics and Autonomous Systems: Synchronizing a swarm of drones or robots can improve task execution, such as in search-and-rescue missions, environmental monitoring, or package delivery, where spatial coordination and communication efficiency are paramount.
%Sensor Networks: In sensor arrays, synchronization can lead to optimized data acquisition and energy efficiency, critical for large-scale deployment in areas like agriculture, pollution monitoring, or disaster management.
%Biological Swarms: Understanding and controlling synchronization within biological swarms can provide deeper insights into animal behavior, aiding in the development of bio-inspired algorithms for engineered systems.
%By investigating synchronization control, we seek to develop a deeper understanding of the governing principles of swarmalators and unlock their full potential in a wide range of applications.

\section{1D swarmalator Model} \label{sec2}
\begin{figure}[t]
	{\centerline{
			\includegraphics[width=\linewidth]{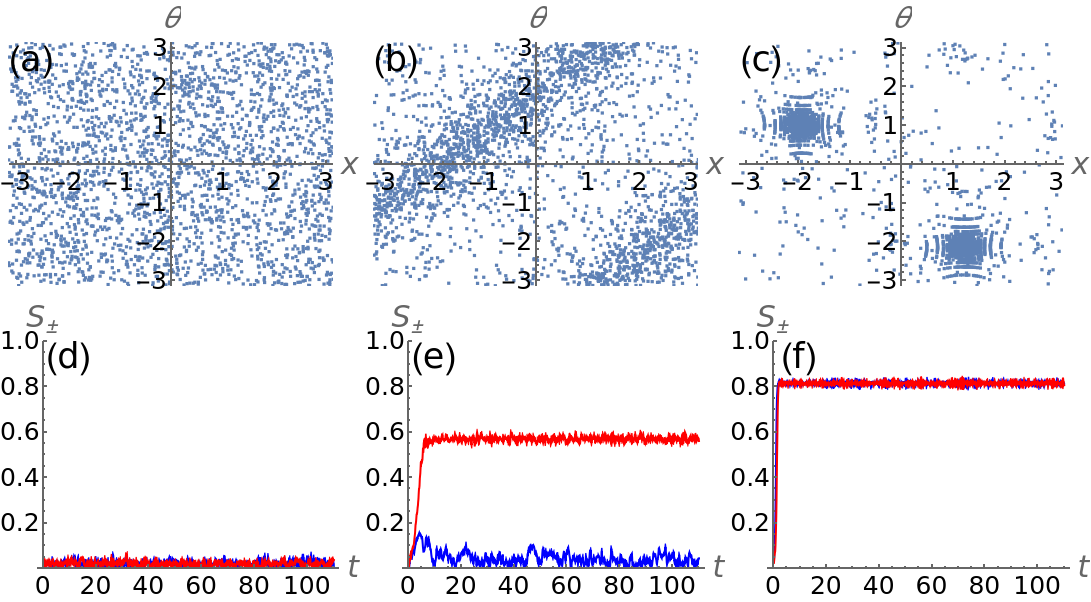}}}
	\caption{Collective states of the 1D swarmalator model. (a) and (d) async state: $(J,K)=(1,1)$. (b) and (e) phase wave state: $(J,K)=(10,1)$. (c) and (f) sync state: $(J,K)=(10,10)$. Simulation parameters are $(t,dt,N)=(100,0.01,2500)$. In the top row, the snapshots of the swarmalators are depicted in the $x$-$\theta$ plane. In the bottom row, blue and red curves stand for the time evolution of $S_+$ and $S_-$, respectively.}
	\label{nonidentical}
\end{figure}

The swarmalator model defined on a ring is given by
\begin{equation} 
\begin{array}{cc}
    \dot{x}_{i}= u_{i}+\dfrac{J}{N} \sum\limits_{j=1}^{N} \sin(x_{j}-x_{i}) \cos(\theta_{j}-\theta_{i})\, ,
\end{array}
\label{sw_position_xeq}
\end{equation}
\begin{equation}  
\begin{array}{cc}
    \dot{\theta}_{i}= \omega_{i}+\dfrac{K}{N} \sum\limits_{j=1}^{N} \sin(\theta_{j}-\theta_{i}) \cos(x_{j}-x_{i}) \, , 
\end{array}
\label{sw_phase_eq}
\end{equation}
where $(x_i,\theta_i)\in \mathbb{S}^1 \times \mathbb{S}^1$ are the positions and phases of the $i$-th swarmalator, $u_i,\omega_i$ are the intrinsic velocities and natural frequencies of the uncoupled swarmalators, respectively, both chosen from Lorentzian distributions with half-widths $\Delta_u$ and $\Delta_w$. $(J,K)$ represent the spatial and phase coupling strengths, $N$ is the total number of swarmalators. The model looks like a pair of Kuramoto models where the spatial dynamics is modulated by the phases and the other way around. To measure the correlation between the spatial positions and phases, the following order parameters are defined
\begin{equation}
    \begin{array}{l}
         W_{\pm}=S_{\pm}e^{\mathrm{i}\psi_{\pm}}=\dfrac{1}{N}\sum\limits_{j=1}^{N}e^{\mathrm{i}(x_{j}\pm\theta_{j})} \, .
    \end{array} \label{op}
\end{equation}
One can also define the so-called  rainbow order parameters~\cite{o2022collective}. In terms of these order parameters the model \eqref{sw_position_xeq}-\eqref{sw_phase_eq} can be rewritten as 
\begin{multline} 
    \dot{x}_{i} = u_{i}+\dfrac{J}{2}\Big(S_{+} \sin(\psi_{+}-(x_{i}+\theta_{i})) \\+ S_{-} \sin(\psi_{-}-(x_{i}-\theta_{i}))\Big)\, ,\label{sw_position_xeq2} 
\end{multline}
\begin{multline}  
    \dot{\theta}_{i} = \omega_{i} +\dfrac{K}{2} \Big(S_{+} \sin(\psi_{+}-(x_{i}+\theta_{i})) \\ - S_{-} \sin(\psi_{-}-(x_{i}-\theta_{i}))\Big) \, .\label{sw_phase_eq2}
\end{multline}
This model exhibits three different collective states and the behaviors of the order parameters allow to distinguish them. When the coupling strengths $J$ and $K$ are small, swarmalators' positions and phases remain incoherent and the {\em asynchronous} (async) state is observed (see Figs.~\ref{nonidentical}(a) and (d)) where both $S_{\pm}$ lie near $0$, corresponding thus to $x_i$ and $\theta_i$ (almost) uniformly distributes in the phase space. As one of the coupling strengths goes beyond the critical value $J+K = 4 (\Delta_u + \Delta_w)$, the positions and phases remain distributed, but they get correlated to each other, giving a nonzero value of $S_+$ or $S_-$. This state, known as the {\em phase wave}, is show in Figs.~\ref{nonidentical}(b) and (e), and one can appreciate the emergence of an (almost) ordered state. Swarmalators completely overcome the disordering effects of their distributed velocities and frequencies and lock their positions and phases when the coupling strengths satisfy the condition $\Delta_u/J + \Delta_w/K < 1/2$. Now, both the order parameters exhibit equal nonzero values, the resulting state is called the {\em synchronous state} (sync) and it is presented in Figs.~\ref{nonidentical}(c) and (f). Other than these three states, an intermediate mixed state is also observed where the order parameters attain nonzero unequal values. We refer the readers to Ref.~\cite{yoon2022sync} for a detailed description and analysis of this model.

\section{Hamiltonian formulation of the swarmalator model} 
\label{sec3}

The control term that we hereby develop relies on the use of Hamiltonian control theory~\cite{vittot2004perturbation} and thus it is rooted on the existence of an Hamiltonian function that embeds the swarmalators dynamics. Let us thus define the $N$-dimensional Hamiltonian system given by
\begin{multline}
	H({\boldsymbol{\theta}}, {\bf I}) = \sum\limits_{i=1}^{N} \omega_{i} I_{i} 
	-\dfrac{K}{N}\sum\limits_{i,j} \sqrt{I_{i}I_{j}} (I_{j}-I_{i})\times \\ \times\cos(x_{j}-x_{i})\sin(\theta_{j}-\theta_{i}), \label{hamiltonian}
\end{multline}
where ${\bf I}=(I_{1}, \dots, I_{N})^\top$ denotes the action variables associated to the phases of the swarmalators $\boldsymbol{\theta} = (\theta_1,\dots,\theta_N)$. It is important to note here that the positions of the swarmalators, $x_i$, are considered as parameters in this setting, it thus results that the interactions among the phases are weighted by the positions. More explicitly, let us consider the phase of the $i$-th swarmalator, $\theta_i$, located at $x_i$, then the phase interaction among the $i$-th and the $j$-th swarmalator is given by $\sin (\theta_j - \theta_i)$ which is weighted by a quantity that depends on their respective positions, i.e., $\cos (x_j - x_i) \equiv W_{ij}$. Observe that $W_{ij}$ is symmetric in its indices. We can thus rewrite the Hamiltonian function~\eqref{hamiltonian} as
\begin{align}
	H({\boldsymbol{\theta}}, {\bf I}) &= \sum\limits_{i=1}^{N} \omega_{i} I_{i} \nonumber \\&
	-\dfrac{K}{N}\sum\limits_{i,j} \sqrt{I_{i}I_{j}} (I_{j}-I_{i}) W_{ij}\sin(\theta_{j}-\theta_{i}) \nonumber \\
	&\equiv H_0({\bf I_{\theta}}) + V({\boldsymbol{\theta}},{\bf I})\, , \label{hamiltonian-1}
\end{align}
where $H_0$ denotes the integrable part representing the uncoupled phase dynamics of the swarmalators. $V$ denotes the nonlinear term arising from the interaction among the swarmalators and can be considered as a perturbation to $H_0$ under the assumption that the coupling strength $K$ is small, i.e., the phases are weakly coupled. The model defined by Eq.~\eqref{hamiltonian-1} is representative of a class of systems that can characterize the Bose-Einstein condensate in a tilted optical lattice and the Lipkin-Meshkov-Glick (LMG) model in the thermodynamic limit \cite{lipkin1965validity,thommen2003classical}.

\par The time evolution of the action and phase variables can be obtained from the Hamiltonian equation as
\begin{multline} \label{}
	\dot{I}_{i} = - \frac{\partial H}{\partial \theta_i} \\ = -2\dfrac{K}{N} \sum\limits_{j=1}^{N} \sqrt{I_{i}I_{j}} (I_{j}-I_{i}) W_{ij} \cos(\theta_{j}-\theta_{i}),
\end{multline}
\begin{multline}  \label{}
	\dot{\theta}_{i}= \frac{\partial H}{\partial I_{i}} =\omega_{i} +\dfrac{K}{N}\sum\limits_{j=1}^{N} \Bigg[2 \sqrt{I_{i}I_{j}} W_{ij} \sin(\theta_{j}-\theta_{i}) \\-\sqrt{\dfrac{I_{\theta_{j}}}{I_{i}}}  (I_{j}-I_{i}) W_{ij} \sin(\theta_{j}-\theta_{i})    \Bigg], 
\end{multline}
for $i=1,\dots, N$. It is noticeable that $I_{i}=c$, for all $i$, are constants of motions for any fixed real $c>0$. Stated differently, the invariant Kuramoto torus defined by $\mathcal{T}^K:= \{({\boldsymbol{\theta}}, {\bf I}) \in \mathbb{S}^N \times \mathbb{R}^N_+ : I_{i}=1/2 \; \;  \forall i\}$, is such that the Hamiltonian flow restricted on the latter coincides with the swarmalators phase dynamics.

\section{Hamiltonian Control of the swarmalators}
\label{sec:hamctrl}
One can prove the existence of an interesting link between the stability of the Kuramoto invariant torus and the synchronization of the phase variables; more precisely if the actions exhibit an unstable behavior close to the Kuramoto invariant torus, then the oscillators achieve a synchronized state~\cite{witthaut2014kuramoto}. Because our goal is to decrease the synchronization of the swarmalators, we will look for a control term to be added to the Hamiltonian function $H({\boldsymbol{\theta}}, {\bf I})$ capable to increase the stability of the Kuramoto invariant torus. Let us observe that we require this term to be small, i.e., smaller than the nonlinear term $V({\boldsymbol{\theta}}, {\bf I})$ in Eq.~\eqref{hamiltonian-1}, and also to be always present in the system. Indeed we are not considering a switched system but an added control term whose intensity increases only when needed; namely once the system is going toward synchronization the control plays a relevant role and act so to decrease the phases synchronization, that in turn reduce the position synchronization.

Vittot and colleagues proposed a method to stabilize the invariant torus by adding a small control term~\cite{vittot2004perturbation,ciraolo2004control}, $f(V) \sim O(V^2)$, to the Hamiltonian $H$, resulting in a controlled Hamiltonian: $H_{ctrl} = H_0 + V + f(V)$. This approach is less invasive than traditional control techniques and allows for quick responses to abnormal dynamics without requiring continuous system measurements, making it a self-organized control strategy. 

The control term $f(V)$ can be derived by using the operator $\{H_0\}$ whose action on the vector space of $\mathcal{C}^{\infty}$ real functions defined on the phase space is given by $\{H_0\}f := \{H_0,f\}$, where $\{\cdot,\cdot\}$ denotes the Poisson bracket. The control term is then formally written as
\begin{equation}
    f(V) = \sum_{n \geq 1} \frac{\{-\Gamma V\}^n}{(n+1)\;!} (n \mathcal{R}V + V) = f_1 + f_2 + \dots\, ,
\end{equation}
where $\Gamma$ and $\mathcal{R}V$ denote the pseudoinverse operator of $H_0$, and the resonant part of $V$, respectively. The series converges under specific conditions, requiring in particular non-resonant frequencies $\omega_i$. By simplifying the resonant part $\mathcal{R} V$, and by assuming a Diophantine condition on $\omega$, the control term ensures the stability of the invariant torus. Even though the full series might not be needed, truncating it to the first term, $f(V) \sim O (V^2)$, can still effectively preserve the Kuramoto torus~\cite{gjata2017using,asllani2018minimally,moriamelucascarletti2024}. This method embeds the Kuramoto model into the Hamiltonian system, enabling control of both through the stabilization of the invariant torus. We refer the interested readers to Ref.~\cite{gjata2017using} for a detailed description of the control mechanism that we hereby adopt.

Through a series of straightforward but lengthy calculations, one can transform the phase evolution by incorporating the computed control term to obtain
\begin{equation}  \label{eq:ctrlswarm}
\begin{array}{cc}
    \dot{\theta}_{i}= \omega_{i}+\dfrac{K}{N} \sum\limits_{j=1}^{N} \sin(\theta_{j}-\theta_{i}) \cos(x_{j}-x_{i}) +\theta_{i}^{C} \,, 
\end{array}
\end{equation}
where the control term $\theta_{i}^{C}$ is given by
\begin{widetext}
\begin{multline} \label{phase_control_term}
    \theta_{i}^{C}= -\dfrac{K^2}{4N^2} \Bigg[
    \sum\limits_{j} \cos(x_{j}-x_{i}) \cos(\theta_{j}-\theta_{i})
    \sum\limits_{l} \dfrac{1}{\omega_{l}-\omega_{i}} \cos(x_{l}-x_{i}) \cos(\theta_{l}-\theta_{i}) \\
    + \sum\limits_{j} \dfrac{1}{\omega_{j}-\omega_{i}} \cos(x_{j}-x_{i}) \sin(\theta_{j}-\theta_{i})
    \sum\limits_{l} \cos(x_{l}-x_{i}) \sin(\theta_{l}-\theta_{i}) \\
    - \sum\limits_{l} \Bigg( \cos(x_{l}-x_{i}) \cos(\theta_{i}-\theta_{l}) 
    \sum\limits_{j} \dfrac{1}{\omega_{j}-\omega_{l}} \cos(x_{l}-x_{j}) \cos(\theta_{j}-\theta_{l}) \\
    + \dfrac{1}{\omega_{i}-\omega_{l}} \cos(x_{l}-x_{i}) \sin(\theta_{i}-\theta_{l}) 
    \sum\limits_{j} \cos(x_{l}-x_{j}) \sin(\theta_{j}-\theta_{l}) 
    \Bigg) \Bigg] \,.
\end{multline}
\end{widetext}
By visual inspection of the latter expression, one can realize that it is smaller than $V$, indeed the control term is of order $K^2$ while $V\sim O(K)$. We can also observe the presence of denominators of the form $\omega_i-\omega_l$, that are well defined because of the non-resonance condition.

\par By introducing two local order parameters that depend on the swarmalator index:
\begin{equation}
    \begin{array}{l}
         \Tilde{W}^{k}_{\pm}= \Tilde{S}^{k}_{\pm}e^{\mathrm{i}\psi^{{k}}_{\pm}}=\dfrac{1}{N} \sum\limits_{j=1}^{N} \dfrac{e^{\mathrm{i}(x_{j} \pm \theta_{j})}}{\omega_{j}-\omega_{k}}\, ,
    \end{array}
\end{equation}
and by using the global order parameters, the control term~\eqref{phase_control_term} can be rewritten as
\begin{widetext}
\begin{multline}\label{ctrl_term_op}
    \theta_{i}^{C}=-\dfrac{K^2}{16} \Bigg[ S_{+}\Tilde{S}^{i}_{+} \cos(\psi_{+}-\psi^{i}_{+}) +S_{+}\Tilde{S}^{i}_{-} \cos(\psi_{+}+\psi^{i}_{-}-2x_{i}) +S_{-}\Tilde{S}^{i}_{+} \cos(\psi_{-}+\psi^{i}_{+}-2x_{i})+S_{-}\Tilde{S}^{i}_{-} \cos(\psi_{-}-\psi^{i}_{-}) -\mathcal{P}_{i}  \Bigg]\,,
\end{multline}
where $\mathcal{P}_{i}$ is defined by
\begin{multline}\label{pk}
    \mathcal{P}_{i}=\dfrac{1}{N} \sum\limits_{l}\Bigg[\bigg(\cos\Big((x_{i}+\theta_{i})-(x_{l}+\theta_{l})\Big) + \cos\Big((x_{i}-\theta_{i})-(x_{l}-\theta_{l})\Big)\bigg) \bigg(\Tilde{S}^{l}_{+}\cos\Big(\psi^{l}_{+}-(x_{l}+\theta_{l})\Big) + \Tilde{S}^{l}_{-}\cos\Big(\psi^{l}_{-}-(x_{l}-\theta_{l})\Big) \bigg) \\ 
    + \dfrac{1}{\omega_{i}-\omega_{l}} \bigg(\sin\Big((x_{i}+\theta_{i})-(x_{l}+\theta_{l})\Big) - \sin\Big((x_{i}-\theta_{i})-(x_{l}-\theta_{l})\Big)\bigg) \bigg(S_{+}\sin\Big(\psi_{+}-(x_{l}+\theta_{l})\Big) - S_{-}\sin\Big(\psi_{-}-(x_{l}-\theta_{l})\Big) \bigg)    \Bigg]\, .
\end{multline}
\end{widetext}

\section{Numerical results}
\label{sec:numres}
The goal of this section is to present the results of dedicated numerical simulations we performed on the controlled swarmalator system to prove the efficiency of first order control term, $\theta_{i}^{C}$, in reducing synchronization both in space and phases. Let us observe that in the following we slightly modify Eq.~\eqref{eq:ctrlswarm} by considering as control term, $\gamma \theta_{i}^{C}$, where the real positive parameter $\gamma$ can be used to tune the strength of control: a zero value indicates no control and nonzero value characterizes a controlled system. 

\subsection{Effectiveness of the control mechanism}
\begin{figure}[t]
	{\centerline{
			\includegraphics[width=1.1\linewidth]{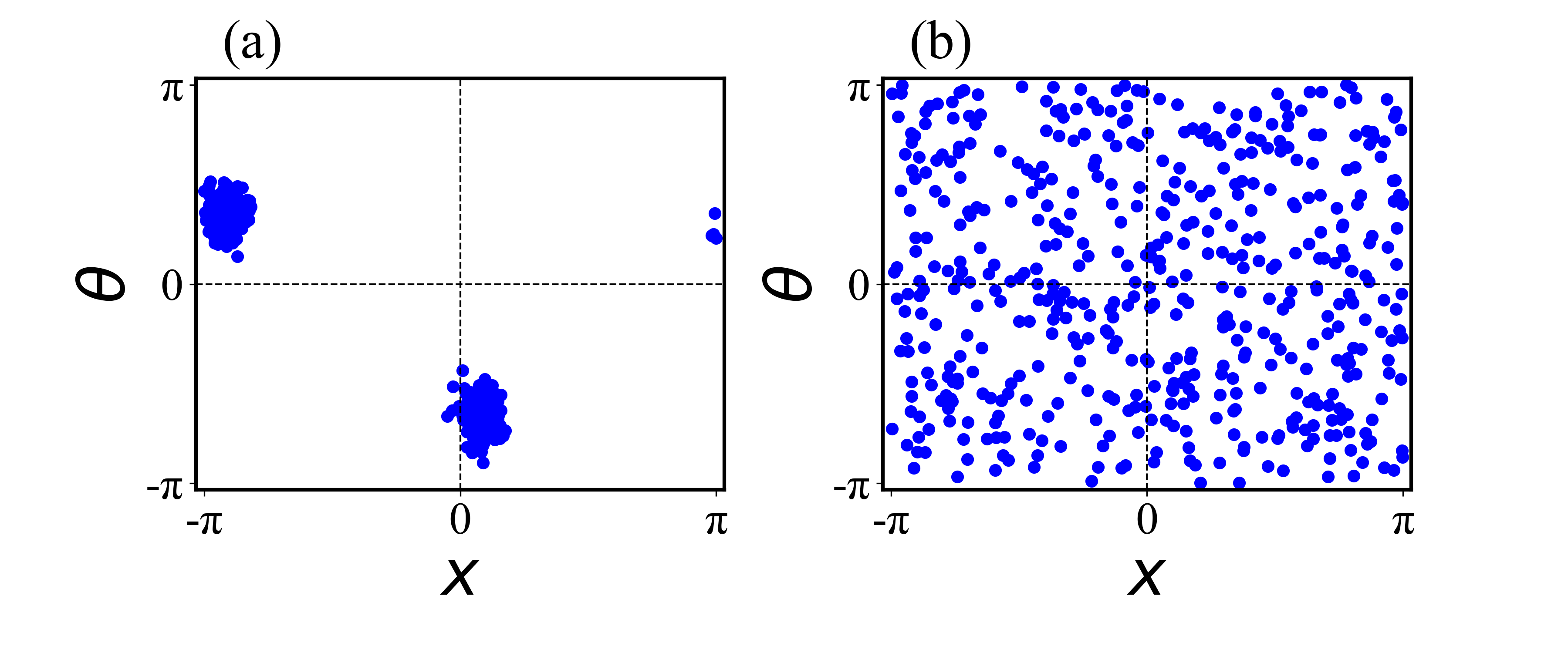}}}
	\caption{Scatter plot in $(x,\theta)$ plane for $(J, K)=(9,5)$. (a) For the uncontrolled system $(\gamma=0)$, swarmalators organize themselves to form the sync state, while in (b) for the controlled system $(\gamma=1)$ they remain in the async state. The results are obtained by integrating the uncontrolled and controlled systems over a time period $t=200$, with the scatter plots displayed at the final time instance.}
	\label{normal_control_uncontrolled_scatter}
\end{figure}

\begin{figure}[t]
	{\centerline{
			\includegraphics[width=1.1\linewidth]{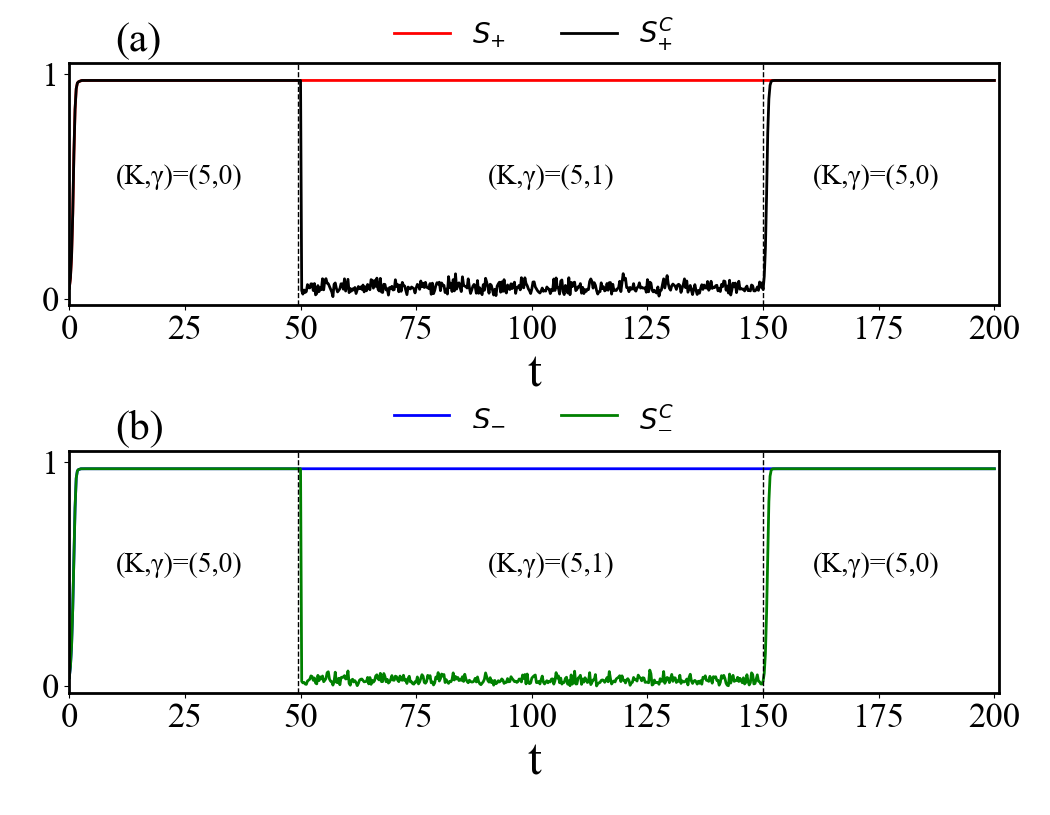}}}
	\caption{Order parameters $S_+$ (top panel, red curve for the uncontrolled population and black curve for the controlled population) and $S_-$ (bottom panel, blue curve corresponding to the uncontrolled population while green curve is used for the controlled one) as a function of time for $N=1000$ swarmalators. Panel (a) displays $S_{+}$, while panel (b) shows the behavior of $S_{-}$. Here, we start the simulation with coupling parameters $(J, K, \gamma)=(9,5,0)$ for which the population of swarmalators exhibit sync state. After $t=50$, we introduce the control term through $\gamma=1$ for up to $t=150$. Beyond this, we again make the system uncontrolled. It can be observed that for the controlled system, the swarmalators exhibit an async state, whereas for the uncontrolled one, it is in the usual sync state.}
	\label{normal_control_on_off}
\end{figure}

Let us recall that our goal is to reduce, or even suppress, the synchronization among the swarmalators both in the spatial and phase components or in any of them by controlling only the phases. So, we start by taking values of the coupling strengths $J$ and $K$ for which the uncontrolled system exhibits the sync state. By assuming that the half-widths of the Lorentzian distributions for the velocity and the frequency are given by $\Delta_u = \Delta_w = 1$, the condition of the stability of the sync state is $(J+K)/JK<1/2$. Based on this bound, we take $J=9$ and $K=5$ to ensure the uncontrolled swarmalators display the sync state with nonzero values of $S_{\pm}$. See the scatter plot in the $x$-$\theta$ plane in Fig.~\ref{normal_control_uncontrolled_scatter}(a). When the control term~\eqref{ctrl_term_op} is considered, we observe that the spatial and phase synchronization are suppressed and swarmalators' positions and phases become desynchronized, as clearly shown  in Fig.~\ref{normal_control_uncontrolled_scatter}(b). This fact is also testified by the values of the two order parameters that now lie near zero. Let us describe the scenario with the help of Fig.~\ref{normal_control_on_off} where the coupling parameters are fixed at $(J,K)=(9,5)$ (see also Supplementary movie~\cite{movieFig3}). In Fig.~\ref{normal_control_on_off}(a) we show the time series behavior of $S_+$ for the uncontrolled (red) and controlled (black) populations. The evolution of $S_-$ is plotted in Fig.~\ref{normal_control_on_off}(b) where blue and green curves correspond to the uncontrolled and controlled populations, respectively. Starting from the initial time $t=0$ the system is effectively uncontrolled (by choosing $\gamma=0$ for the controlled system) up to time $t=50$ and both order parameters correctly lie near $1$. At time $t=50$, the control term is turned on (by making $\gamma=1$) and its effect is realized instantly as the order parameters of the controlled system ($S^C_{\pm}$) go to zero. The control is kept activated till $t=150$ and it is observed that in the time interval $[50,150]$ the order parameters of the controlled system stay near zero indicating the async state. Then control is turned off at $t=150$ (by making $\gamma=0$) and the system returns to the sync state as the order parameters reacquire their values close to $1$. This observation reveals that the control term successfully tackles the synchronization among swarmalators' positions and phases and it is able to bring incoherence in the system that we aimed for.

\begin{figure}[htp]
    \centerline{
    \includegraphics[width=1.0\linewidth]{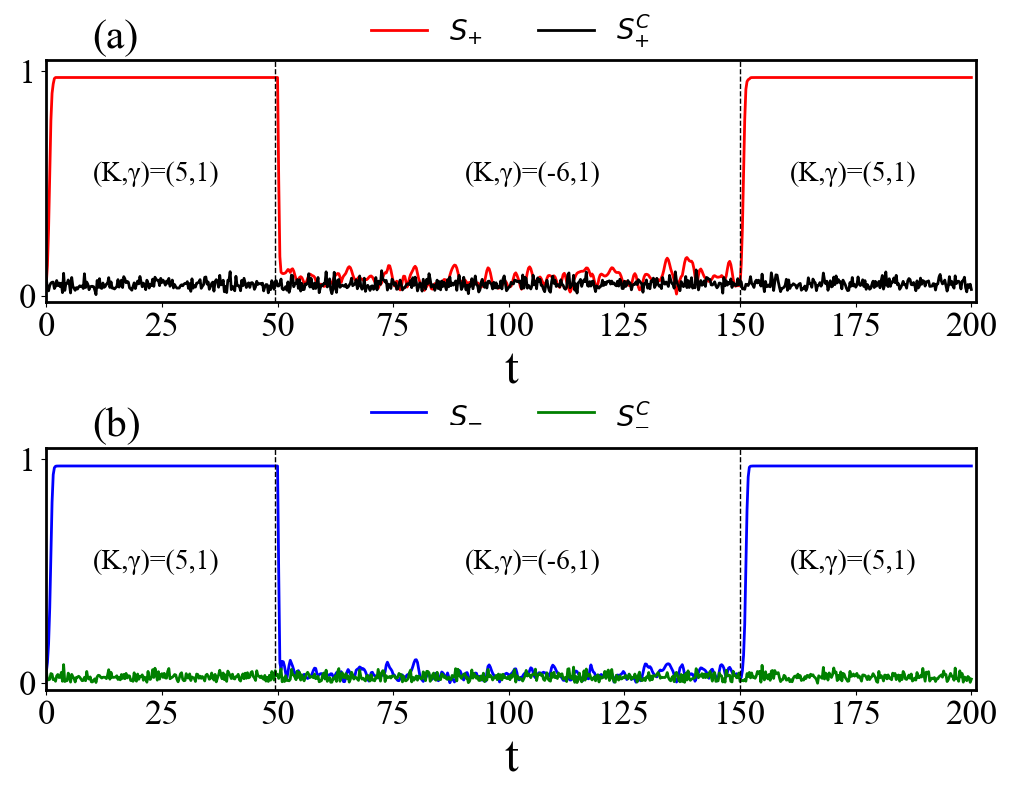}}
    \caption{Panel (a) displays $S_{+}$, while panel (b) corresponds to $S_{-}$. At time $t=0$ we set the coupling parameters $(J, K)=(9,5)$ for which the population of swarmalators exhibits the sync state for uncontrolled $(\gamma=0)$ system and async for controlled $(\gamma=1)$ systems. In the time interval  $[50,150]$, we decrease the coupling parameter $K$ to $-6$, value for which both the uncontrolled and controlled swarmalators remain in the async state. After $t=150$ we increase the coupling parameter back to $K=5$, for which the controlled system remains in the async state while the uncontrolled system achieves the sync state. Let us also observe that in the async region, the controlled and the original system show similar behavior, because the control term does not affect the system much while it is only effective for the region of interest, i.e., the region of sync state.}
    \label{normal_op_time_vary_kp_sync}
\end{figure}

We also show the effectiveness of the control by varying the coupling parameter after starting from the sync state. In Fig.~\ref{normal_op_time_vary_kp_sync}, we start by choosing coupling parameters from the sync region by taking $(J,K)=(9,5)$. The order parameters of the uncontrolled system (red and blue curves) lie near $1$ and signify the sync state. The controlled system's order parameters (black and green curves) lie near zero that highlights the async state. At time $t=50$, phase coupling parameter is changed to $K=-6$ which now belongs to the async region. One can find that the order parameters of both the controlled and the uncontrolled systems remain close to zero, i.e., both the systems exhibit the async state. This pattern exists till $t=150$ where $K$ is brought back to its starting value $K=5$. The uncontrolled system spontaneously goes back to the sync state, whereas the controlled one remains in the async state. See the caption of Fig.~\ref{normal_op_time_vary_kp_sync} and Supplementary movie~\cite{movieFig4} for more details.

\begin{figure}[htp]
    \centerline{
    \includegraphics[width=1.1\linewidth]{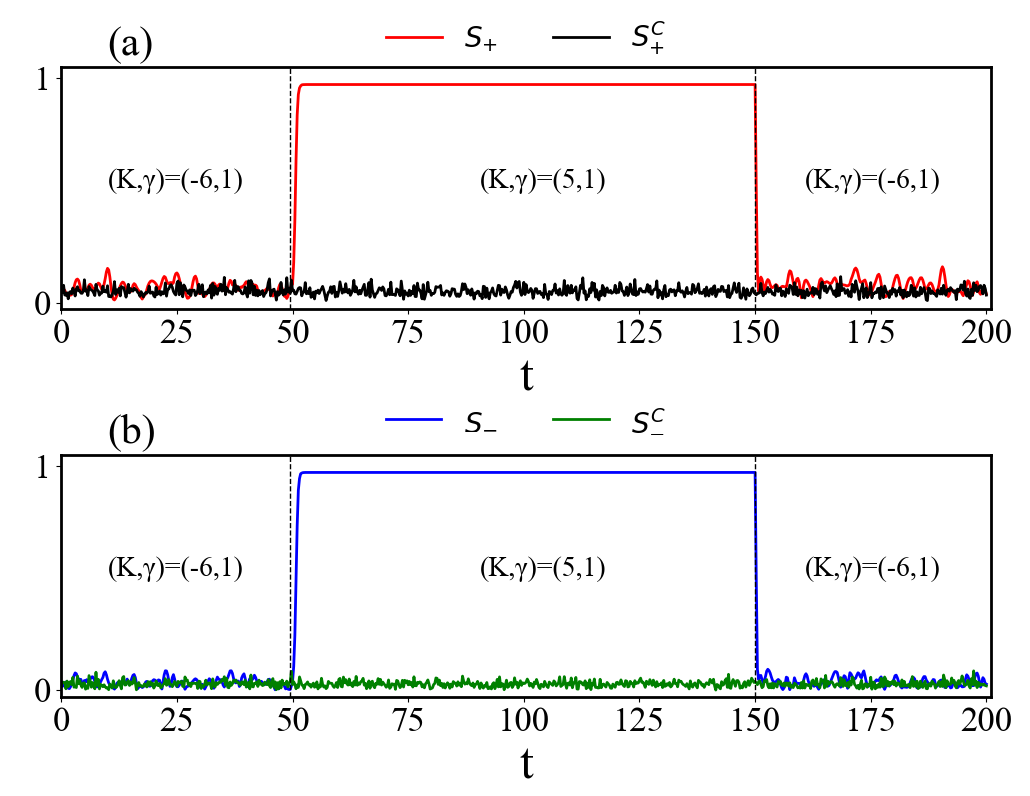}}
    \caption{Panel (a) displays $S_{+}$ and panel (b) corresponds to $S_{-}$. At time $t=0$ we set the coupling parameters $(J, K)=(9,-6)$ for which the population of swarmalators exhibits the async state for both uncontrolled $(\gamma=0)$ and controlled $(\gamma=1)$ systems. In the time interval  $[50,150]$, we increase the coupling parameter $K$ to $5$, value for which the uncontrolled swarmalators move into the sync state, on the other hand the controlled system remains in the async state. After $t=150$ we decrease the coupling parameter back to $K=-6$ and both systems lie again in the async state. }
    \label{normal_op_time_vary_kp}
\end{figure}

Let now consider a complementary case where, i.e., the system is initialized in the async state, then after some time we modify the coupling parameters to let the uncontrolled system to move to the sync state while the controlled one will remain in the async state. To achieve this goal we  choose $J=9$ and $K=-6$ that satisfy the condition, $J+K<4$, for the uncontrolled system to lie in the async state. By looking at Fig.~\ref{normal_op_time_vary_kp}(a,b) we can realize that the rainbow order parameters for the uncontrolled ($\gamma=0$) and the controlled ($\gamma=1$) swarmalators, are small, certifying thus the async state, and moreover very similar each other. Let us indeed observe that because the control term is designed to reduce the synchronization, when the system is already in the async state, the dynamics is not modified by the control term. At time $t=50$, we increase the phase coupling strength to $K=5$ which now correspond to the sync region. As anticipated, we can see that $S_{\pm}$ promptly deviate from zero and go close to $1$, i.e., the uncontrolled swarmalators now exhibit the sync state. On the other hand, the controlled system can subdue the effect of the change in $K$ and it remains in the async state as one can appreciate by the fact that $S^C_{\pm}$ stay close to $0$. This behavior holds true in the time interval $[50,150]$, then at $t=150$ we set $K$ back to the initial value $-6$, starting from this point both the systems find themselves in the async state. The takeaway of this analysis is that the proposed control strategy is capable of inhibiting the change of the coupling strength and holds the system to the async state (see also Supplementary movie~\cite{movieFig5}).

\subsection{Robustness to variations of the coupling strength}
\begin{figure}[htp]
    \centerline{
    \includegraphics[width=1.1\linewidth]{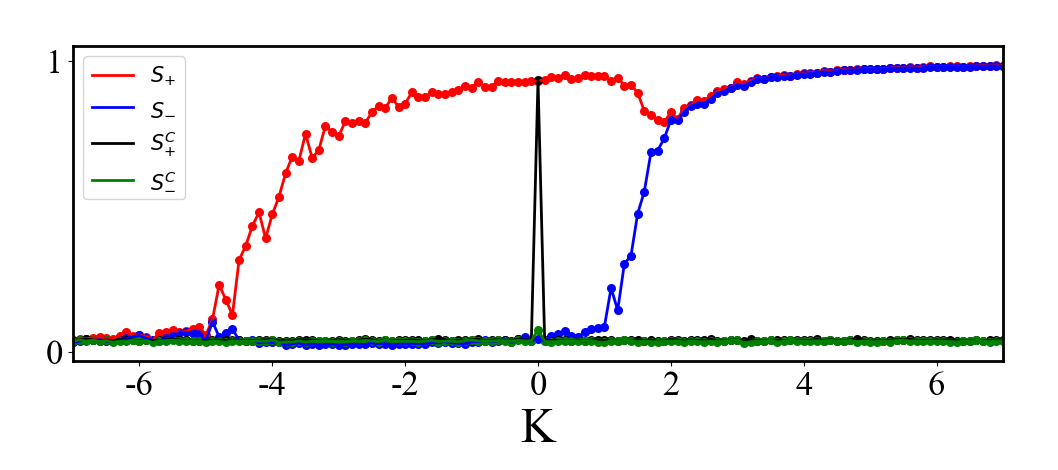}}
    \caption{Order parameters $S_{\pm}$ (colored as before) as a function of phase coupling strength $K$ for $N=1000$ swarmalators. The spatial coupling strength $J$ is kept fixed at $J=9$. For the controlled system we set $\gamma=1$. At each value of $K$, we plot the time-averaged order parameters after discarding a sufficient transient period. Precisely, we integrate the controlled and uncontrolled systems over a time period  $t=200$ and compute the time average of the order parameters using the final $10\%$ of the data. We observe that for the uncontrolled system $(\gamma=0)$, the system goes through the transitions ``async $\xrightarrow{}$ phase wave $\xrightarrow{}$ mixed state $\xrightarrow{}$ sync" with increasing $K$. However, for the controlled system, the population of swarmalators remains in the async state for all the values except $K=0$, where the controlled term is ineffective. Thus, the proposed controlled term shows a significant impact not only in the region of sync state but also in all other states. }
    \label{normal_op_kp} 
\end{figure}

Let us now study the robustness of the control strategy by varying the phase coupling strength $K$ over a range $[-7,7]$, the spatial coupling strength $J$ being fixed to $9$. The uncontrolled swarmalators go through multiple phase transitions as $K$ is varied (see the red and blue curves in Fig.~\ref{normal_op_kp}). Starting at $K=-7$ the async state is found with $S_{\pm} \approx 0$, this state loses its stability and the phase wave state is observed when $K$ is increased to $-5$. Here, $S_+$ deviates from zero while $S_-$ remains at zero. Around $K=2$ both $S_+$ and $S_-$ acquire nonzero equal values that denote the emergence of the sync state. In between, i.e., $K\in [1,2]$, we also observe the presence of the mixed state in a small interval  where $S_+$ and $S_-$ are both nonzero but $S_- < S_+$. In conclusion, we find that the uncontrolled swarmalators go through a series of phase transitions and all the collective states are realized by varying $K$. We now consider the controlled system where we set $\gamma=1$. The order parameters for the controlled system ($S^C_{\pm}$) are plotted with black and green curves in Fig.~\ref{normal_op_kp} and we can realize that they always stay close to $0$ which indicates that the controlled swarmalators always facilitate the async state as we have claimed. The only exception takes place at the particular value $K=0$ where the controlled system goes to the phase wave state ($S^C_+ \approx1, S^C_-\approx 0$). The reason behind this phenomenon is that the control term $\theta^C_i$ given by Eq.~\eqref{ctrl_term_op}  depends on the coupling strength $K$ and when $K=0$ we effectively cancel out the control term; the dynamics of the uncontrolled swarmalators are observed, which is the phase wave in this case.

\subsection{Minimally invasive strategy}
\begin{figure}[htp]
    \centerline{
    \includegraphics[width=1.1\linewidth]{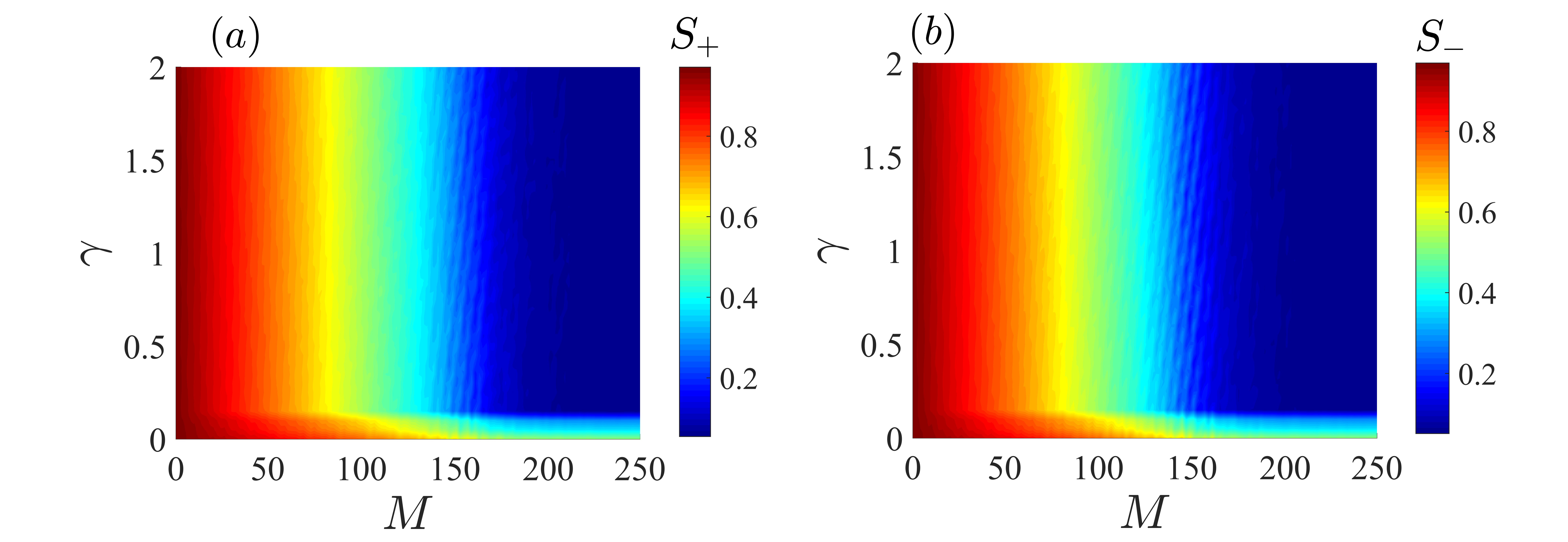}}
    \caption{The swarmalator order parameters $S_{\pm}$ is shown as a function of number of controlled swarmalators $M$ and the strength of the control $\gamma$. The order parameters at each $(M,\gamma)$ are determined by averaging over an extended period, excluding initial transients. Panel (a) shows the variations of $S_{+}$, while panel (b) corresponds to $S_{-}$. For each pair $(M,\gamma) \in [1,250] \times [0,30]$, we numerically integrate the system with $N=250$ and coupling strengths $(J,K)=(9,5)$. It is observed that it is sufficient to only control a fraction of swarmalators to achieve a required state and moreover the result is not dependent on the control strength $\gamma$ after a critical value $\gamma \approx 0.2$.}
    \label{normal_m_gamma}
\end{figure}

In the previous sections we have shown that the control strategy is very efficient in reducing, or even suppressing, the synchronization of the swarmalators. In this section we will prove that the strategy is also minimally invasive, namely we do not need to control all the swarmalators to desynchronize them, one can achieve a similar result by acting on a fraction of them. Let us thus assume to control $M$ swarmalators out of the total population $N$, namely the function $\theta_i^C$ is added only to $M$ indexes $i$ (that we assume to be $\{1,\dots,M\}$), while the dynamics of remaining ones is ruled by the original swarmalator system~\eqref{sw_phase_eq}. Under this assumption, the local order parameters become
\begin{equation}
	\begin{array}{l}
		\Hat{W}^{k}_{\pm}= \Hat{S}^{k}_{\pm}e^{\mathrm{i}\Hat{\psi}^{{k}}_{\pm}}=\dfrac{1}{M} \sum\limits_{j=1}^{M} \dfrac{e^{\mathrm{i}(x_{j} \pm \theta_{j})}}{\omega_{j}-\omega_{k}} ,
	\end{array}
\end{equation}
for $k=1,2,\dots,M$. The restricted control term is now achieved by substituting $\Tilde{S}^i_{\pm}$ by $\Hat{S}^i_{\pm}$ in Eq.~\eqref{ctrl_term_op}. To check the capability of the simplified control to reduce the synchronization, we perform a series of numerical experiments by varying $M$ and the strength of the control, $\gamma$, and we compute the values of the order parameters $S_{\pm}$. As previously done, we choose the coupling strengths ensuring the original model to lie in the sync region, i.e., $(J,K)=(9,5)$. The results are shown in Fig.~\ref{normal_m_gamma} where we report the order parameters $S_{\pm}$ in the case of a system made of $N=250$ swarmalator. One can observe that once the control strength $\gamma$ is larger than a critical threshold ($\approx 0.2$), controlling a fraction of the whole population ($M\approx 125$) is sufficient for successfully bringing the swarmalators to the async state, corresponding to the blue regions where $S_{\pm}$ is close to $0$. Let us observe that this fraction corresponds to (almost) half of the total population and thus support the claim that the control strategy is not very invasive. Beyond $\gamma =0.2$, the behavior remains independent of the strength of the control.

\subsection{Truncated control term}  
The aim of this section is to propose a simplified control term still capable of reducing the synchronization. The starting point is the full control term~\eqref{ctrl_term_op} and we focus on the term denoted by $\mathcal{P}_{i}$. The latter is computationally much more expensive compared to the remaining terms in the expression of $\theta^C_i$. As a result we define the truncated control term $\theta_i^{tr}=\theta^C_i-\mathcal{P}_{i}$, where, i.e., we neglected the contribution from  $\mathcal{P}_{i}$. Our aim is now to show that this new control term is still able to reduce synchronization. What we observe here is very interesting and different from the previously studied scenarios. Indeed, by starting from the sync state, the addition of the truncated control term with a small value of the control strength ably makes the the swarmalators desynchronized both in their positions and phase, but the positions and phases remain correlated to each other. Saying differently, from the sync state one reaches the phase wave state where one of $S_+$ and $S_-$ is greater than $0$ while the other is almost $0$. With a larger value of the control strength $\gamma$, the correlation between the phases and positions also perishes and both $S_+$ and $S_-$ go to $0$. We infer that the control strength also plays a role here in desynchronizing the swarmalators.
\begin{figure}[t]
    \centerline{
    \includegraphics[width=1.1\linewidth]{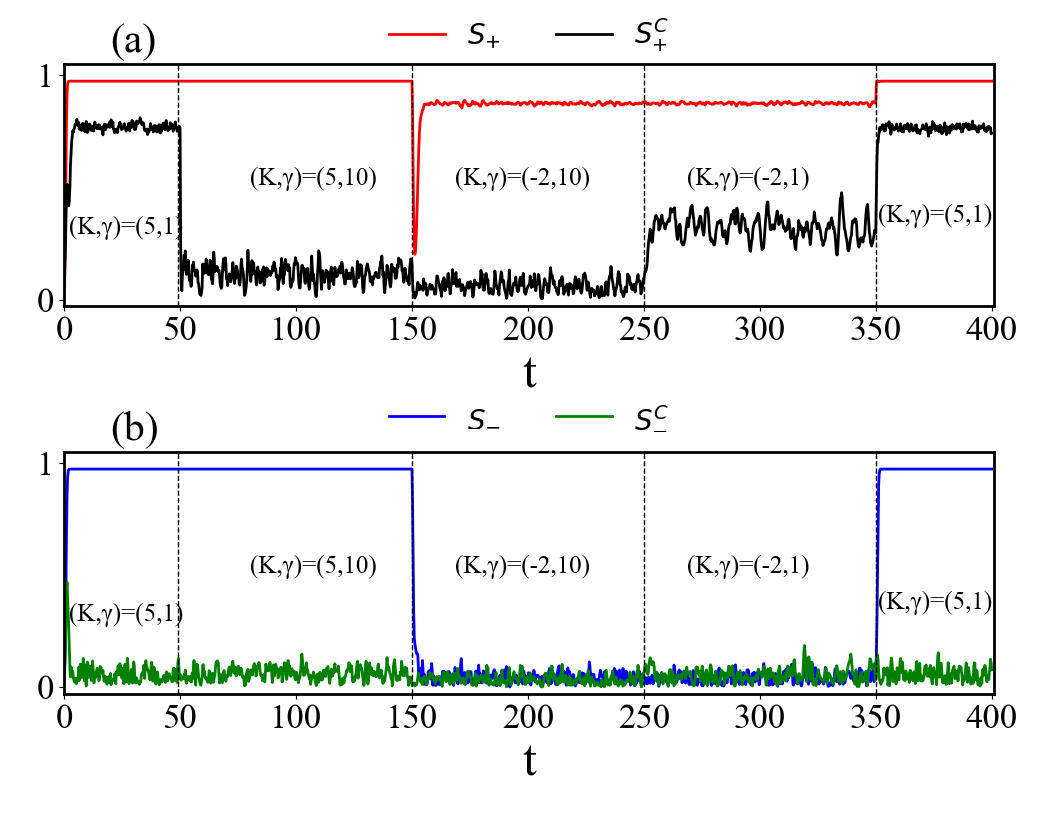}}
    \caption{Order parameters $S_{\pm}$ as a function of time for $N=1000$ swarmalators. Panel (a) displays $S_{+}$, while panel (b) corresponds to $S_{-}$. Here, we start the simulation with coupling parameters $(J, K)=(9,5)$ and control strength $\gamma=1$ for the controlled population of swarmalators. At $t=50$, we increase the control strength to $\gamma=10$ until $t=150$. For $t \in [150,250]$, $\gamma$ remains at $10$ while $K$ is reduced to $-2$. From $t=250$ to $t=350$, $\gamma$ is lowered back to $1$. Beyond this, $K$ is increased again to $5$ while keeping $\gamma$ fixed at $1$. It can be observed that the synchronized population exhibits a phase wave state at lower control strengths, while a higher control strength leads to an asynchronous state.}
    \label{truncated_op_time_vary_kpc}
\end{figure}

Figure~\ref{truncated_op_time_vary_kpc} demonstrates this scenario (see also Supplementary movie~\cite{movieFig8}). The coupling strengths are chosen from the sync region of the uncontrolled system as $(J,K)=(9,5)$. The time series of the order parameters $S_+$ and $S_-$ are plotted in Figs.~\ref{truncated_op_time_vary_kpc}(a) and (b), respectively for both the uncontrolled and controlled populations with the same color scheme used in the previous figures. Initially, $S_+$ and $S_-$ lie near $1$ as the system exhibits the sync state. For the controlled system, $\gamma$ is fixed to $1$ from $t=0$ till time $t=50$, and in this interval $S^{C}_-$ goes to zero but $S^{C}_+$ remain nonzero, i.e., the controlled swarmalators exhibit the phase wave state (first segment from left in Figs.~\ref{truncated_op_time_vary_kpc}(a,b)). At $t=50$ we increase the control strength to $\gamma=10$ and observe that $S^{C}_+$ also goes to zero. The controlled population is now fully desynchronized in both their positions and phases and are void of any correlation between them (second segment). Then at $t=150$ we make a change in the phase coupling strength by reducing it to $-2$ from the initial value $K=5$. This value of $K$ belongs to the phase wave region, and as expected one of $S_+$ and $S_-$ drops to zero from $1$ while the other remains nonzero. The controlled system continues to stay at the async state as the control strength is still kept at $\gamma=10$ (third segment). Next at $t=250$, we reduce the control strength to $1$ from $10$. This value of the control strength is no longer sufficient to keep the swarmalators at the desynchronized state and we see that $S^{C}_+$ increases from zero to some nonnegative value, i.e., the controlled population also goes to the phase wave state (fourth segment). Finally at $t=350$, the coupling strength is increased back to $5$ from $-2$. We regain the behavior that we observed in the first segment for both the controlled and uncontrolled populations.

%The pattern continues until we further increase the control strength at $t=150$ by choosing a larger value 10. It is readily observed that now ($S^{C}_+$) also decays to 0 along with ($S^{C}_-$) already being close to 0. That means swarmalators have reached the async state where their positions and phases are incoherent and are void of any correlation between them. The scenario persists till $t=250$ where we again decrease $\gamma$ back to 1 and the system returns to the phase wave state. Finally, at $t=350$ we turn off the control by making $\gamma$ to be 0 and instantly the sync state is realized where we started from.

\begin{figure}[hpt]
    \centerline{
    \includegraphics[width=1.1\linewidth]{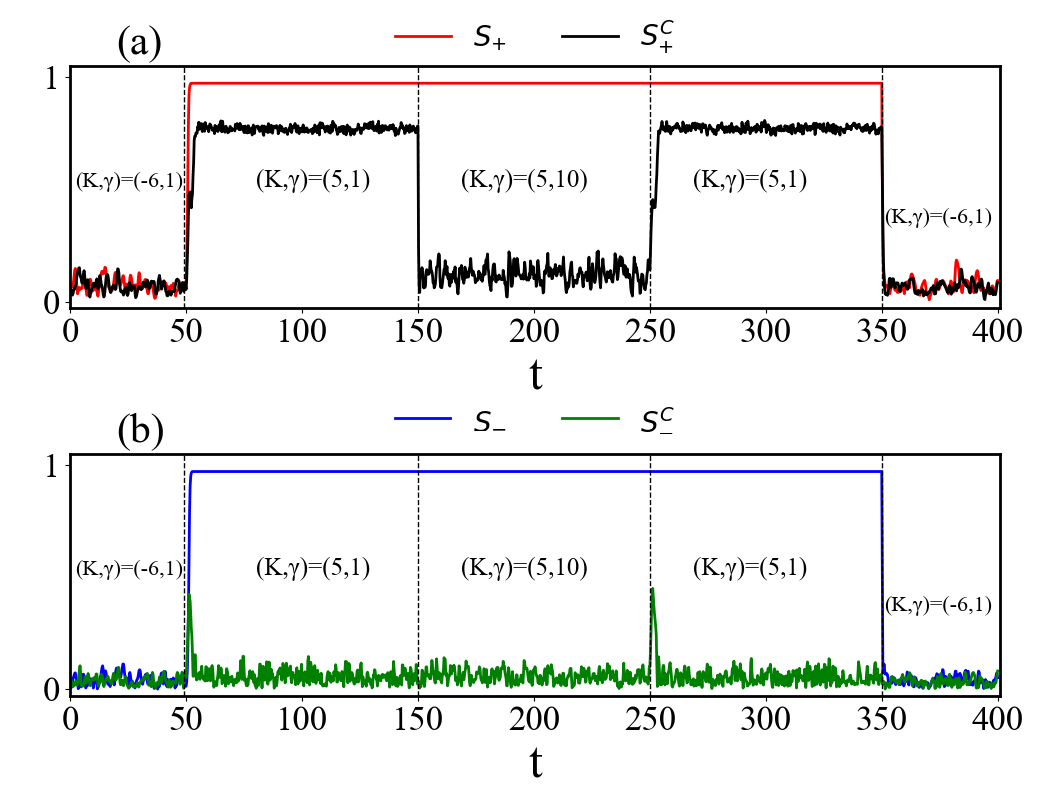}}
    \caption{Order parameters $S_{\pm}$ as a function of time for $N=1000$ swarmalators. Panel (a) displays $S_{+}$, while panel (b) corresponds to $S_{-}$. Here, we start the simulation with coupling parameters $(J, K)=(9,-6)$ for which the population of swarmalators exhibits an async state for both uncontrolled $(\gamma=0)$ and controlled $(\gamma=1)$ systems. After $t=50$, we increase the coupling parameter $K$ to $5$ for up to $t=150$. We increase $\gamma$ to $10$, keeping $K$ fixed at $K=5$ for $t \in [150,250]$. Thereafter, we decrease the strength of control back to $\gamma=1$. Beyond this, we again decrease it back to $K=-6$, while $\gamma$ is kept fixed at $\gamma=1$. It can be observed that for the controlled system, the swarmalators exhibit the phase wave and async state depending on the strength of control, whereas, for the uncontrolled one, it goes to the sync state when the coupling strength is increased. It is also observable that in the region of the async state, the control term does not affect the system much and is only effective for the region of interest, i.e., the region of the sync state. }
    \label{truncated_op_time_vary_kp}
\end{figure}

Next, we perform the complementary analysis by starting from the async region of the uncontrolled system by choosing $(J,K)=(9,-6)$. The controlled system also belongs to the async state in that case (first segment in Figs.~\ref{truncated_op_time_vary_kp}(a,b)). Once we increase the coupling strength to $K=5$, the uncontrolled system goes to the sync state. But, we find that a comparatively smaller value of the control strength ($\gamma=1$) is unable to curb the effect of the increase in $K$ and the controlled system goes to the phase wave state from the async state (second segment). With a larger value of $\gamma=10$, full desynchronization is achieved (third segment). It is not possible to sustain this complete asynchrony by reducing $\gamma$ which can be seen from the fourth segment where the controlled system goes to the phase wave state from the preceding async state once we decrease $\gamma$ from $10$ to $1$. When we reduce $K$ to $-6$ both systems, controlled and uncontrolled one, go to the async state (last segment)  (see also Supplementary movie~\cite{movieFig9}).

\begin{figure}[hpt]
    \centerline{
    \includegraphics[width=1.1\linewidth]{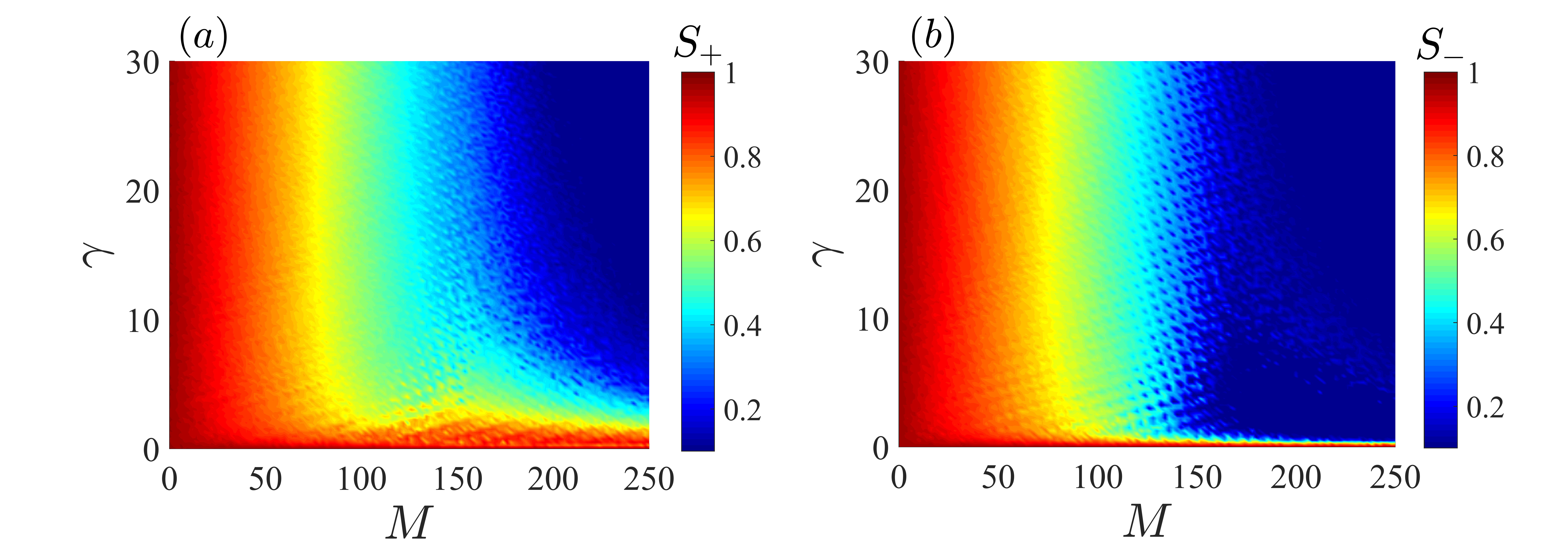}}
    \caption{The swarmalator order parameters $S_{\pm}$ as a function of number of controlled swarmalators $(M)$ and the strength of the control $\gamma$. Panel (a) shows the variations of $S_{+}$, while panel (b) corresponds to $S_{-}$. For each pair $(M,\gamma) \in [1,250] \times [0,30]$, we numerically integrate the system with $N=250$ and coupling strengths $(J,K)=(9,5)$. It is observed that only a sufficient number of swarmalators can be controlled to achieve a required state, and it is dependent on the control strength $\gamma$.}
    \label{trucated_m_gamma}
\end{figure}

The main observation is that here we can also reach a phase wave by tuning on $\gamma$ and $M$, which was not possible with the full control term. We can make the control strategy minimally invasive by not only controlling a fraction of the entire population but also by tuning the control strength. In Fig.~\ref{trucated_m_gamma} we show the results obtained by varying these quantities and investigate the change in the order parameters $S_{\pm}$ when the coupling strengths are fixed in such a way the uncontrolled system lies in the sync region. The main observation is that only a fraction of the population is needed to be controlled to achieve  desynchrony ($S_{\pm}\approx0$), moreover the fraction depends on the control strength: the larger $\gamma$ the smaller is the number of required controllers. Let us also observe the presence of a minimal number of controllers, say $M\sim N/2$, required to substantially reduce the synchronization even for very large $\gamma$.

\section{Discussions}
In this paper, we have introduced a Hamiltonian control strategy to tackle the synchronization in the 1D swarmalator model. Swarmalators are entities that self-organize their positions and synchronize their phases. For the 1D swarmalator model, both positions and phases are periodic variables that belong to $\mathbb{S}^1$ and the dynamics look like a pair of entangled Kuramoto models. With large spatial and phase coupling strengths, both the positions and phases get synchronized, our goal is to reduce, or even suppress, this synchronized phenomenon by acting on the phases only. To achieve this goal, we first introduced a suitable Hamiltonian function whose dynamics reduce to the one of the swarmalator of a given invariant torus. Then, relying on the link between stability of the invariant torus and synchronization, we build a Hamiltonian control term capable to increase the stability of the invariant torus~\cite{witthaut2014kuramoto}. We observe that the control term curbs synchronization irrespective of the value of the control strength, provided it is nonzero. Also, only a fraction of the swarmalators is needed to be controlled to achieve a reduction of synchronization. When the smaller amplitude terms are neglected and removed from the control term, we find that the control strength becomes relevant. A smaller control strength, although desynchronizes the swarmalators, cannot overcome the correlation between the positions and phases. A larger value of the control strength is required to establish complete desynchrony among swarmalators where the correlation between the positions and phases also vanishes.

In this study, we have obtained and studied a control strategy based on a Hamiltonian function where only the phases have been taken into account, while the positions have been considered ``as parameters'' in the Hamiltonian function, and thus we have only controlled phases of the swarmalators. We remark that one could have used a more general Hamiltonian function depending on both phases and spatial positions and thus build a complete control term where both variables are controlled. However, the control terms become very large, computationally too expensive and it is thus not clear the added value with respect to the effective control strategy proposed in this work; a more detailed studied is thus left for a future study.

\bibliographystyle{apsrev}
\bibliography{ref}

\end{document}